%% file: 00-main-sigconf.tex
\documentclass[sigconf,natbib=true,anonymous=false]{acmart}

\usepackage{enumitem}
\usepackage{epstopdf}
\usepackage{multirow}
\usepackage{balance}
\usepackage[ruled,linesnumbered]{algorithm2e}
\usepackage{caption}
\usepackage{subcaption}
\usepackage{graphicx}
\usepackage{float}
\usepackage{adjustbox}
\usepackage{wrapfig,lipsum}
\usepackage{booktabs,etoolbox}
\usepackage{amsmath,amsfonts,amsthm,bm}
\usepackage{layouts}
\usepackage{cleveref}
\usepackage{tablefootnote}
\usepackage{makecell}
\usepackage{upgreek}
\usepackage{mathtools, stmaryrd}
\usepackage{algpseudocode}
\usepackage{xparse}

\AtBeginDocument{%
  \providecommand\BibTeX{{%
    \normalfont B\kern-0.5em{\scshape i\kern-0.25em b}\kern-0.8em\TeX}}}

\copyrightyear{2023} 
\acmYear{2023} 
\setcopyright{acmlicensed}\acmConference[SIGIR '23]{Proceedings of the 46th International ACM SIGIR Conference on Research and Development in Information Retrieval}{July 23--27, 2023}{Taipei, Taiwan}
\acmBooktitle{Proceedings of the 46th International ACM SIGIR Conference on Research and Development in Information Retrieval (SIGIR '23), July 23--27, 2023, Taipei, Taiwan}
\acmPrice{15.00}
\acmDOI{10.1145/3539618.3591769}
\acmISBN{978-1-4503-9408-6/23/07}

\usepackage{xspace}
\newcommand{\MLIR}{MLIR\xspace}

\begin{document}

\title{Soft Prompt Decoding for Multilingual Dense Retrieval}

\author{Zhiqi Huang}
\email{zhiqihuang@cs.umass.edu}
\affiliation{%
  \institution{University of Massachusetts Amherst, USA}
  \city{}
  \state{}
  \country{}
}

\author{Hansi Zeng}
\email{hzeng@cs.umass.edu}
\affiliation{%
  \institution{University of Massachusetts Amherst, USA}
  \city{}
  \state{}
  \country{}
}

\author{Hamed Zamani}
\email{zamani@cs.umass.edu}
\affiliation{%
  \institution{University of Massachusetts Amherst, USA}
  \city{}
  \state{}
  \country{}
}

\author{James Allan}
\email{allan@cs.umass.edu}
\affiliation{%
  \institution{University of Massachusetts Amherst, USA}
  \city{}
  \state{}
  \country{}
}

\renewcommand{\shortauthors}{Zhiqi Huang, Hansi Zeng, Hamed Zamani, \& James Allan}

\begin{abstract}
In this work, we explore a Multilingual Information Retrieval (MLIR) task, where the collection includes documents in multiple languages.
We demonstrate that applying state-of-the-art approaches developed for cross-lingual information retrieval to MLIR tasks leads to sub-optimal performance. This is due to the heterogeneous and imbalanced nature of multilingual collections -- some languages are better represented in the collection and some benefit from large-scale training data. To address this issue, we present KD-SPD, a novel soft prompt decoding approach for MLIR that implicitly ``translates'' the representation of documents in different languages into the same embedding space. 
To address the challenges of data scarcity and imbalance, we introduce a knowledge distillation strategy. The teacher model is trained on rich English retrieval data, and by leveraging bi-text data, our distillation framework transfers its retrieval knowledge to the multilingual document encoder.
Therefore, our approach does not require any multilingual retrieval training data.
Extensive experiments on three MLIR datasets with a total of 15 languages demonstrate that KD-SPD significantly outperforms competitive baselines in all cases. We conduct extensive analyses to show that our method has less language bias and better zero-shot transfer ability towards new languages.

\end{abstract}



\maketitle
\input{sections/introduction}

\input{sections/related_work}

\input{sections/methodology}
\input{sections/experiments}
\input{sections/conclusion}

\begin{acks}
This work was supported in part by the Center for Intelligent Information Retrieval and This research is based upon work supported in part by the Center for Intelligent Information Retrieval, and in part by the Office of the Director of National Intelligence (ODNI), Intelligence Advanced Research Projects Activity (IARPA), via Contract No. 2019-19051600007 under Univ. of Southern California subcontract no. 124338456. The views and conclusions contained herein are those of the authors and should not be interpreted as necessarily representing the official policies, either expressed or implied, of ODNI, IARPA, or the U.S. Government. The U.S. Government is authorized to reproduce and distribute reprints for governmental purposes notwithstanding any copyright annotation therein.. Any opinions, findings and conclusions or recommendations expressed in this material are those of the authors and do not necessarily reflect those of the sponsor.
\end{acks}

\bibliographystyle{ACM-Reference-Format}
\balance
\bibliography{01-reference}
\end{document}

%% file: sections/introduction.tex
\section{Introduction}  \label{sec:introduction}

In Cross-Lingual Information Retrieval (CLIR), a user submits a query in one language, and the system responds by retrieving documents in another language. Hence, in addition to a ranking model, CLIR systems require an extra component of language translation to align the vocabulary in the query language with that of the document language. 
This translation gap can be bridged by employing dictionaries~\cite{yarmohammadi-etal-2019-robust}, statistical translation tables \cite{bonab2020training}, machine translations \cite{sarwar-etal-2019-multi}, 
or, more recently, multilingual pre-trained large language models~\cite{huang2021mixed}. 

Motivated by many real-world applications, such as web search, where the retrieval collection includes documents from multiple languages \cite{chaware2009information, kishida2005technical}, this work focuses on a multilingual retrieval setting, where the query is in one language and the collection is a mixture of languages. We refer to this task as \MLIR, and it has been previously explored by \cite{lin2002description, peters2012multilingual, rahimi2020axiomatic}. 
Even though CLIR and \MLIR are tightly coupled, effective \MLIR models must overcome additional major challenges. For instance, instead of one pair of languages between query and document, the translation component in the \MLIR model needs translation knowledge for multiple language pairs. \citet{xu2001evaluating} found that in such scenarios, the distribution of relevant documents to a given query often differs in different languages -- which highlights the challenges in designing effective \MLIR models that also perform fairly across languages. 


The advent of multilingual versions of pre-trained Transformer-based language models, such as mBERT~\cite{devlin2018bert} and XLM-R~\cite{conneau2019unsupervised}, provides the possibility of jointly learning representations for many languages. Because tokens in different languages are projected into the same semantic space, the pre-training phase imparts the model with multilingual translation knowledge.
Like monolingual retrieval, fine-tuning these models with multilingual retrieval data allows the model to learn the knowledge of query-document matching and perform retrieval tasks under a multilingual setting.

\begin{figure}[t]
    \captionsetup[subfigure]{font=footnotesize,labelfont=footnotesize}
    \begin{subfigure}[t]{0.4\textwidth}
        \centering
        \includegraphics[width=0.9\linewidth]{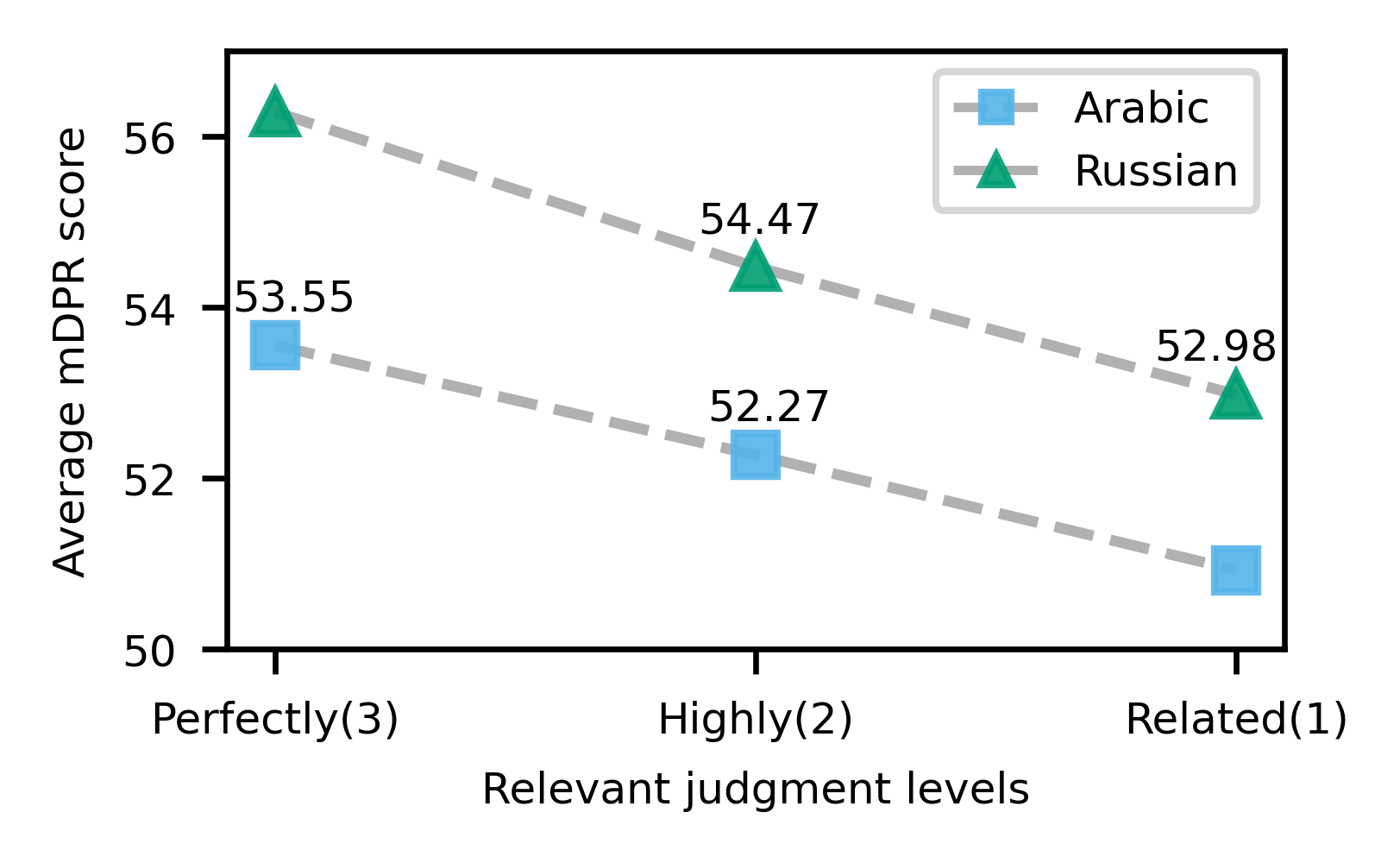}
    \end{subfigure}
\caption{Average score given to parallel documents in Arabic and Russian by mDPR \cite{zhang2021mr}. Queries and relevant judgments are from the TREC 2020 Deep Learning Track. Passages are translated by mMARCO~\cite{bonifacio2021mmarco}}
\label{fig:language-bias}
\end{figure}
However, we find that this modeling pipeline, which delivers state-of-the-art results in many CLIR tasks \cite{galuvsvcakova2021cross, yu2020study, yang2022c3},
suffers from two major shortcomings when applied to \MLIR settings. First, to learn query-document matching knowledge on multiple language pairs effectively, this models requires access to multilingual \emph{retrieval} training data that covers the languages present in the target collection. However, many languages suffer from the scarcity of multilingual training data with reliable relevance judgment~\cite{lawrie2022hc4}. Therefore, it is challenging to achieve broad language coverage in training data. For languages not covered in the training data, the model has to retrieve documents in a so-called zero-shot manner, creating a performance gap between the observed and non-observed languages~\cite{macavaney2020teaching}. 
Second, due to the unbalanced pre-training data in different languages, the performance of multilingual pre-trained models varies by language in many downstream tasks~\cite{wu2020all, wang-etal-2020-extending}. \MLIR models built on such pre-trained models can inherit language bias, leading to inconsistent ranking results. To demonstrate this case, we pair the test queries from TREC 2020 Deep Learning Track~\cite{craswell2020overview} with their relevant passages translated into Arabic and Russian by mMARCO~\cite{bonifacio2021mmarco}. Then for each language, we score query-document pairs using the multilingual dense passage retriever (mDPR)~\cite{zhang2021mr}. Figure~\ref{fig:language-bias} illustrates the difference in ranking the same set of relevant documents in these two languages. We observe that mDPR scores Russian documents higher than their Arabic version. We argue that such language bias in \MLIR would lead to sub-optimal ranking results, e.g., highly relevant documents in Arabic have lower scores than slightly relevant documents in Russian. 

To address these issues, we present KD-SPD,\footnote{KD refers to the model training framework, and SPD refers to the model architecture.} a multilingual dense retrieval model based on knowledge distillation (KD) and soft prompt decoder (SPD) for the \MLIR task. KD-SPD does not require any multilingual relevance labels for training, thus automatically solving the data scarcity issue in low-resource languages. Our approach solely requires monolingual retrieval training data in English, which we obtain from MS MARCO~\cite{nguyen2016ms}, and a large collection of parallel and comparable documents. Note that such data is abundant and easily collected through automatic bi-text mining algorithms~\cite{elkishky_ccaligned_2020}. We use CCAligned~\cite{elkishky_ccaligned_2020} in our experiments.

We first train a monolingual dense retrieval model $M$, such as ANCE \cite{xiong2020approximate}, for the English language. Since this model has the relevance matching ability, we freeze its document encoder and then minimize the distance between the representations learned by $M$ for any English document and the representations learned by KD-SPD for its parallel or comparable version in other languages. In fact, $M$ acts as a monolingual teacher model for the multilingual student model KD-SPD. Therefore, our approach implicitly ``translates'' the representation of documents in different languages into the same language embedding space. 
We hypothesize that although different languages possess unique properties such as distinct grammar or vocabulary, they also have common traits for expressing similar meanings. 
To capture these unique and shared features, KD-SPD uses decomposable soft prompt, which is derived as the product of a shared matrix and a low-rank language-specific matrix for each language.
Our proposed encoder-decoder architecture transforms documents into contextualized token embeddings and decodes the outputs with language-specific prompts to obtain a final representation.
Through joint training across multiple languages, we observe that the learned prompts are capable of reducing language bias and possess the transferable capacity to generalize to unseen languages. 

We performed extensive experiments on three \MLIR datasets with a total of 15 languages from diverse linguistic families, including both high- and low-resource languages.  We also conduct experiments on different relevant distributions with respect to language.
In terms of mean average precision (MAP), our proposed method significantly outperforms several strong baseline methods in all multilingual settings, including a 20.2\% improvement over mDPR and a 9.6\% improvement over a multilingual knowledge distillation method from Sentence-BERT (SBERT)~\cite{reimers2020making}.
Further analysis demonstrates that KD-SPD has less language bias and better zero-shot transfer ability toward new languages.

%% file: sections/related_work.tex
\section{Related Work}  \label{sec:related_work}
\subsection{Neural Matching Models for \MLIR}
With respect to language settings, \MLIR and CLIR are closely related. CLIR mostly focuses on retrieval between two particular languages, while \MLIR considers multiple language pairs between query and document. 
In general, information retrieval involving a multilingual setting has two sub-tasks: translation and query-document matching. One method involves translating the query into the language of the document set, then using a monolingual retrieval model to evaluate relevance. 
The translation sub-task can be performed using Statistical Machine Translation (SMT)~\cite{bonab2020training} or Neural Machine Translation (NMT)~\cite{saleh-pecina-2020-document}.
The two-step process of translation followed by retrieval is widely used; however, with the advent of bilingual word representation~\cite{vulic2015monolingual, artetxe2017learning} and multilingual pre-trained language models~\cite{devlin-etal-2019-bert, conneau-etal-2020-unsupervised}, it is possible to bypass the translation step and match the query and document in different languages within a shared representation space.

Multilingual pre-trained language models usually prepend a special token to the input sequence to support downstream applications.
Because the special token embedding is contextualized based on other tokens in the sequence, once finetuned, they are effective across various tasks, including retrieval tasks~\cite{litschko2018unsupervised, yu2021cross, Litschko2018, yu2020study}.
Named cross-encoder, the model takes the concatenation of the query and document as input. An embedding of the ``[CLS]'' token is fed into a feed-forward layer to produce a score for the input pair~\cite{nogueira2019passage}. With multilingual knowledge from pre-training, these language models help bridge the vocabulary between query and document languages. Like monolingual retrieval, multilingual retrieval models based on cross-encoder are computationally expensive and usually rely on a lexical-based sparse retrieval as the first step to finding relevant information. 
Dense retrieval based on a bi-encoder architecture is proposed to overcome the sparse retrieval bottleneck~\cite{luan2021sparse, lee2019latent, karpukhin2020dense}. With the separation of the query document encoders, dense retrieval has already shown success on monolingual retrieval tasks~\cite{khattab2020colbert,gao2021unsupervised}. By replacing the underlying language model with its multilingual version, dense retrieval is extended to a multilingual setting~\cite{zhang2021mr}. 

However, the translation gap prevents the multilingual retrieval models from achieving the same level of performance as models in the monolingual (i.e., English-to-English) setting~\cite{huang2021mixed}. 
Supporting the model with abundant multilingual retrieval data is one way to reduce the effect of the translation gap. \citet{sasaki2018cross} constructed large-scale, weakly supervised CLIR collections based on the linked foreign language articles from Wikipedia pages.
\citet{bonifacio2021mmarco} built \MLIR training data using neural machine translation models. Besides retrieval data, approaches like utilizing external knowledge in language-specific modules are also suggested to close the language gap. 
\citet{bonab2020training} showed that when fine-tuned with retrieval data, dictionary-oriented word embedding could improve the performance of a CLIR model. \citet{huang2021mixed} proposed a mixed attention transformer architecture to learn relevance judgments and word-level translation knowledge jointly. 
\citet{yang2022parameter} designed a language adapter component to efficiently transfer models based on monolingual data to a cross-lingual setting. 

These approaches mostly focus on improving CLIR performance where the query and the target documents are from two particular languages. In this work, we focus on MLIR, a more general setting where the document collection comprises a diverse mix of languages, which is gaining increasing attention recently~\cite{li-etal-2022-learning-cross}. 
While being able to bridge the translation gap between multiple languages, the model for \MLIR task also needs to be language-agnostic when ranking documents in different languages. 
Our approach implicitly “translates” documents in different languages into an embedding space tuned for English retrieval.

\subsection{Multi-task \& Prompt-based Learning}

The goal of multi-task learning is to leverage the shared underlying structure of different tasks to improve the performance of each task~\cite{Ruder2017AnOO}. A common approach is to transfer the knowledge from a model fine-tuned on multiple source tasks to the target task \cite{Raffel2019ExploringTL, Vu2020ExploringAP, Aghajanyan2021MuppetMM}. For example, \citet{Aghajanyan2021MuppetMM} introduce a pre-finetuning stage that involves multi-task learning steps on diverse NLP tasks. They show that training stability can be improved by applying task-heterogenous batches with task-rebalancing loss scaling. Recent works show that the zero-shot and few-shot performance of pre-trained large language models can be boosted by prompted multi-task learning \cite{Sanh2021MultitaskPT, Wang2022BenchmarkingGV, Liu2022FewShotPF}. For instance, \citet{Sanh2021MultitaskPT} develop a system that maps any NLP task into a human-readable prompt form where each supervised dataset contains multiple prompts with diverse wording. The experiments imply that the multi-task training on these prompted datasets can improve the zero-shot performance of the pre-trained models. 
Other works \cite{Zhong2021AdaptingLM} focus on zero-shot classification (ZAC), introducing a meta-tuning training paradigm to optimize the zero-shot classification objective via fine-tuning. They consolidate various classification tasks into a single QA format, compiling a dataset of classification tasks with human-authored prompts for meta-tuning.

Soft Prompt tuning has shown great potential to adapt large language models to downstream tasks \cite{Lester2021ThePO, Vu2021SPoTBF, Asai2022ATTEMPTPM, anonymous2023multitask}. \citet{Vu2021SPoTBF} further study the generalizability and transferability of the soft prompts. They first learn a prompt on one or more source tasks and use it as the initialized prompt for a target task. The simple target prompt initialization method can match or outperform full fine-tuning across all model sizes. 
\citet{Asai2022ATTEMPTPM} extend the work by training an attention module to interpolate the source prompts and newly initialized target prompt for each downstream task. During the multi-task training, only the target prompt and attention weights are updated, while the soft prompts and original language model’s parameters are frozen. A recent approach \cite{anonymous2023multitask} learns a transferable shared prompt by applying matrix decomposition and knowledge distillation from multiple source task-specific prompts and using the low-ranking matrix updating for target task adaption.

KD-SPD builds upon the idea of prompt-oriented, parameter-efficient multi-task learning. It treats retrieval in each language as a distinct task while jointly modeling them to capture shared underlying structures. The primary insight is that languages, despite unique properties, share common features and concepts. We utilize decomposable prompts to model these aspects. Unlike conventional parameter-efficient approaches, experiments show that updating prompts jointly with model parameters enhances retrieval performance.


\subsection{Knowledge Distillation}

Proposed by \citet{hinton2015distilling}, knowledge distillation is a method to train a model, called the student, using valuable information provided by the output of another model, called the teacher. This way, the teacher model's knowledge can be transferred into the student model. The idea of knowledge distillation is wildly used in the field of computer vision~\cite{xie2020self, yuan2019ckd, Lin_2022_CVPR}, natural language processing~\cite{sanh2019distilbert, reimers2020making} and information retrieval~\cite{lu2020twinbert, hofstatter2020improving, Santhanam2021ColBERTv2EA, Zeng2022CurriculumLF, lin-etal-2021-batch}.

In the field of information retrieval, it is common for the teacher model to be a complex reranker model with higher capacity but lower efficiency compared to the efficient dual-encoder based student model. \citet{Santhanam2021ColBERTv2EA} apply the KL divergence loss to align query-document scores between teacher and student models. Another approach is balanced topic-aware query sampling~\cite{hofstatter2020improving}, which shows further improvement on top of the original knowledge distillation loss. To address the performance gap, \citet{Zeng2022CurriculumLF} propose a curriculum learning based knowledge distillation framework that trains a student model with increasing difficulty. In addition to monolingual retrieval, multilingual distillation frameworks have also been proposed. \citet{li-etal-2022-learning-cross} explore using query-document scores as the distillation signals. The cross-lingual token alignment task has also been studied as an optimal transport problem, with \citet{Huang2022ImprovingCI} proposing a distillation framework to build a CLIR model via bitext data.

Our model training framework is also an extension of knowledge distillation. 
A typical framework for knowledge distillation relies on a teacher model to solely provide target distributions~\cite{gou2021knowledge,ma2022deep}. 
Our approach has different sources of knowledge: the major knowledge is from the teacher model, and we also consider the cross-lingual knowledge shared by the prompt matrix. Moreover, from the language perspective, rather than focusing on one CLIR task, our model simultaneously learns retrieval knowledge for multiple CLIR tasks.

%% file: sections/methodology.tex
\section{Methodology}  \label{sec:methodology}
Our goal is to incorporate the knowledge of query-document matching from a well-trained monolingual retrieval model into a multilingual transformer-based retrieval architecture, such that it is capable of generating contextual representations under the \MLIR setting and thus performing query-document matching in different languages.
In this section, we first define the \MLIR task and outline our approach. Then we present the key component of our model: a soft prompt-based encoder-decoder architecture. Finally, we introduce the model training via a knowledge distillation framework and build the \MLIR model with components from both the teacher and student models. Due to space limitations, we focus on the \MLIR case of searching a multilingual collection with an English query as an example to describe our method. 
It is worth noting that English may also be included in the multiple collection.

\subsection{Overview}
Given a query $q$ in language $X$ and a target collection $\mathcal{D}_{\bm{Y}}$ which contains documents in language set $\bm{Y} = \{Y_1, Y_2, \ldots Y_K \}$, suppose $d_{ki}$---the $i$\textsuperscript{th} document in language $Y_k$---has the ground truth relevance label $Rel(q, d_{ki})$, then the aim is to design an \MLIR model $f$ that retrieves a list of documents from $\mathcal{D}_{\bm{Y}}$ such that
\begin{equation}\label{eq:mlir}
    f(q, d_{ki}) \geq f(q, d_{lj}), \quad \forall \ Rel(q, d_{ki}) \geq Rel(q, d_{lj})
\end{equation}
where $f(\cdot, \cdot)$ indicates the ranking score calculated by the model.
To build model $f$, we first assume there exists an oracle model $g$ for the retrieval task in language $X$. Thus, given $q$ and monolingual collection $\mathcal{D}_X$, $g$ satisfies:
\begin{equation}
    g(q, d_{xi}) \geq g(q, d_{xj}), \quad \forall \ Rel(q, d_{xi}) \geq Rel(q, d_{xj})
\end{equation}
We can achieve (\ref{eq:mlir}) with model $f'$ if for any $d_*$ in $\bm{Y}$ and its translation $d_x$ in $X$, the model matches the oracle:
\begin{equation*}
    f'(q, d_*) = g(q, d_x)
\end{equation*}
Suppose both $f'$ and $g$ follow the architecture of dense retrieval, the ranking score calculation is the dot-product of the query and document embeddings, thus:
\begin{equation*}
    f'_{E}(q) f'_{D}(d_*)^\top = g_{E}(q) g_{D}(d_x)^\top
\end{equation*}
where $f'_{E}$ and $g_{E}$ are query encoders; $f'_{D}$ and $g_{D}$ are document encoders for $f'$ and $g$ respectively. We then reuse $g_{E}$ as the query encoder of $f'$. With $f'_{E} = g_{E}$, we have:
\begin{equation}\label{eq:zero_distance}
    g_{E}(q)\bigl(f'_{D}(d_*) - g_{D}(d_x)\bigr)^\top = 0
\end{equation}
It is safe to assume $g_{E}(q)$ is a nonzero vector. Therefore the goal of finding $f'$ is equivalent to reducing the embedding distance between parallel documents. In our method, we retrain $g_D$ as the teacher model by removing its parameters from the computational graph and train $f'_D$ as the student model.
Note that in practice, the oracle model $g$ does not exist. We can use an off-the-shelf English-to-English (monolingual) dense retrieval model as a substitute for $g$. 
Because $g_D$ is fixed, the essence of knowledge distillation training is to push multilingual document representations generated by $f'_D$ toward their corresponding English document representations generated by $g_D$. Moreover, Equation \eqref{eq:zero_distance} suggests that the training of $f'_D$ does not rely on either query $q$ or ground truth relevant judgment. A group of parallel or comparable sentences from English to any other language involved in the collection is adequate to train $f'_D$. 
Parallel or comparable sentences between two languages are often referred as bitext data.
Unlike multilingual retrieval data, which often require relevance labels, bitext data are easier to acquire, especially for low-resource languages~\cite{heffernan2022bitext,tan2022bitext}.

\begin{figure}[t]
    \captionsetup[subfigure]{font=footnotesize,labelfont=footnotesize}
    \begin{subfigure}[t]{0.23\textwidth}
        \centering
        \includegraphics[width=\linewidth]{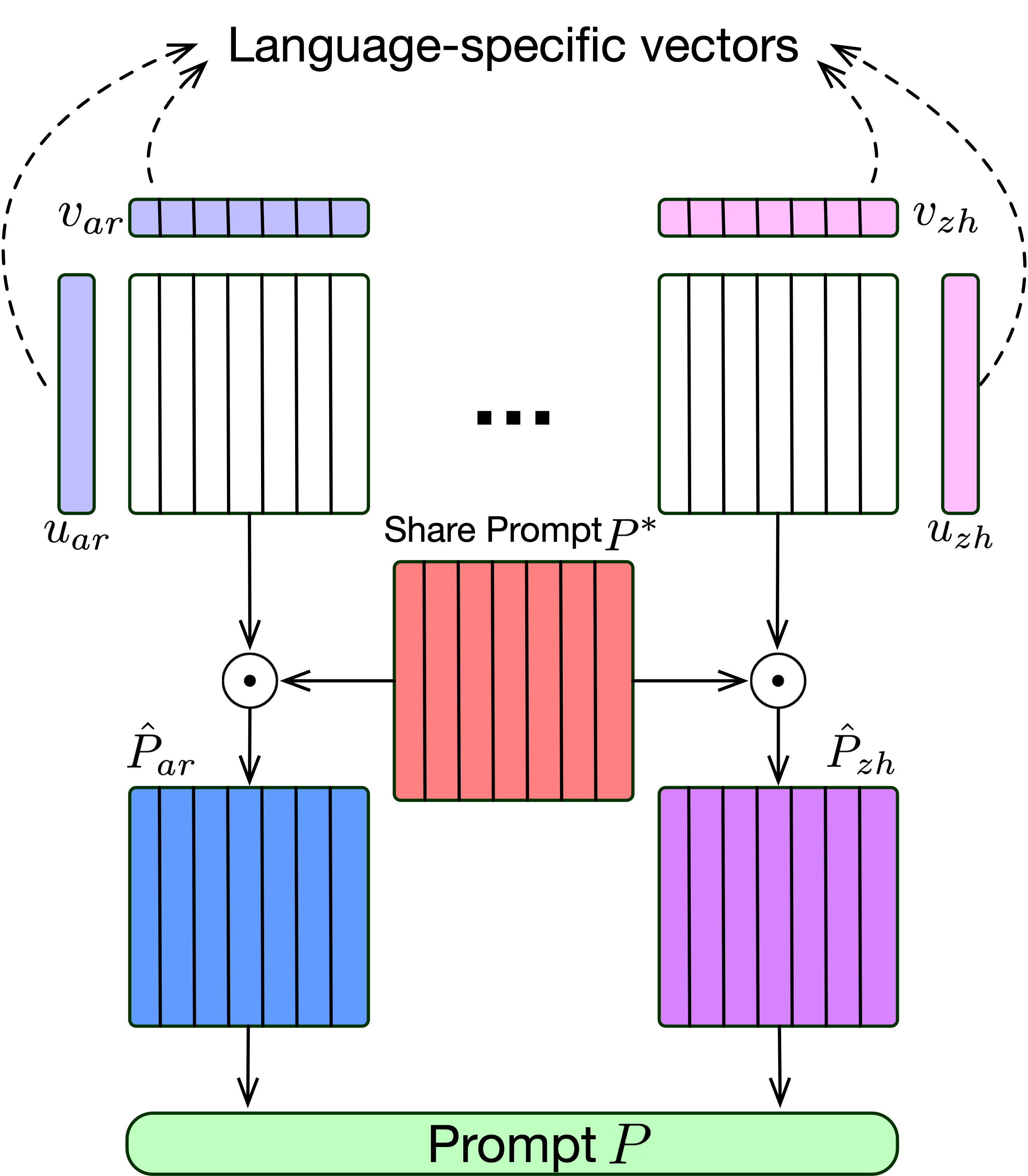}
        \caption{Soft prompt decomposition.}
        \label{fig:prompt}
    \end{subfigure}
    \begin{subfigure}[t]{0.23\textwidth}
        \centering
        \includegraphics[width=\linewidth]{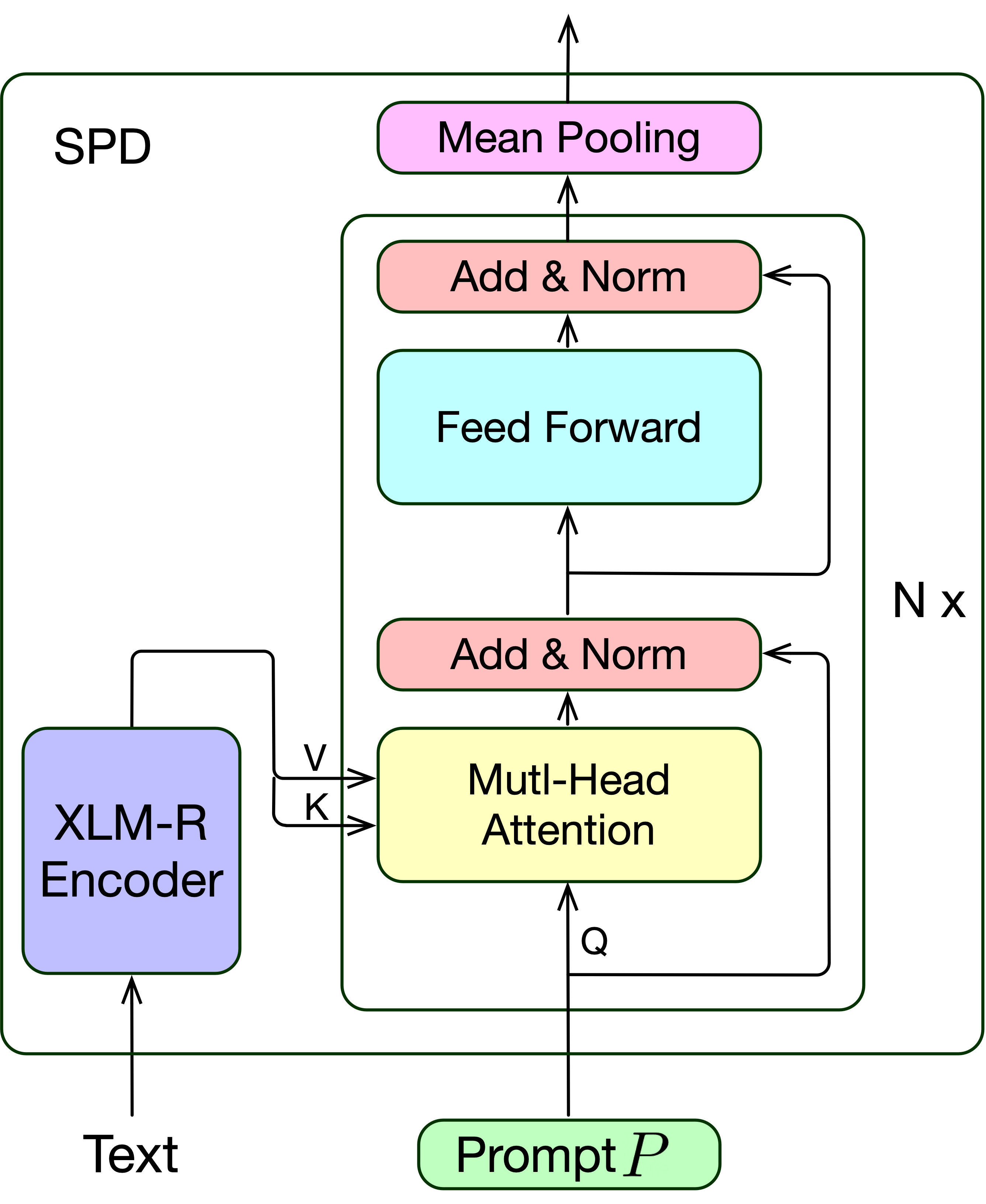}
        \caption{Multilingual document encoder.}
        \label{fig:decoder}
    \end{subfigure}
\caption{SPD model architecture.}
\label{fig:model}
\end{figure}

\subsection{Soft Prompt Decoder}
We focus on the design of the document encoder of the student model, $f'_D$, which handles multilingual documents. In general, the function of $f'_D$ is similar to a neural machine translation model. The difference is that $f'_D$ translates the input text into an embedding in the target language rather than natural language text. Thus, we build $f'_D$ based on the encoder-decoder architecture.
For the encoder component of $f'_D$, we exploit multilingual pre-trained language models (i.e., mBERT or XLM-R). The token representation generated by the encoder is then fed to the decoder component. However, unlike the decoder with an autoregressive generation process, we propose a soft prompt-based decoder (SPD) architecture. 

\noindent \textbf{Soft Prompt Matrix.} 
We consider $f'_D$ as a multitask model where translating (mapping) each language in the multilingual collection into the target language space is viewed as a single task. 
Using the language name as the task identifier, a prompt $\mathbf{P}_k \in \mathbb{R}^{l\times d}$ for language $Y_k$ with the same dimension as the token embedding $d$ and vector length as $l$ is used as input to the decoder. Thus, the prompt matrix serves as the language-based decoding initialization vector. 
Inspired by the prompt decomposition from multitask prompt tuning~\cite{anonymous2023multitask}, we decompose $\mathbf{P}_k$ into two parts, as shown in Figure~\ref{fig:prompt}: language-specific low-rank vectors $\mathbf{u}_k \in \mathbb{R}^{l}$ and $\mathbf{v}_k\in \mathbb{R}^{d}$ for language $Y_k$; And a shared prompt $\mathbf{P}^* \in \mathbb{R}^{l\times d}$ across all languages. 
The language-specific prompt can be parameterized as $\mathbf{W}_k = \mathbf{u}_k \cdot \mathbf{v}_{k}^{\top}$, which has the same dimension as the shared prompt $\mathbf{P}^*$.
The final prompt $\mathbf{\hat{P}}$ for language $Y_k$ is then formulated as follows.
\begin{equation} \label{eq:prompt}
    \mathbf{\hat{P}}_k = \mathbf{P}^* \odot \mathbf{W}_k = \mathbf{P}^* \odot (\mathbf{u}_k \cdot \mathbf{v}_{k}^{\top})
\end{equation}
where $\odot$ denotes the Hadamard product between two matrices. 
The shared prompt enables efficient knowledge sharing across all source languages and commonalities across translation tasks. Meanwhile, the language-specific vectors still allow each translation task to maintain its own parameters to encode language-specific knowledge. 
Additionally, prior studies on multitask prompt learning also showed that soft prompt learned from multitask data can be efficiently transferred to a new task~\cite{vu2021spot, su2022transferability}. In section~\ref{sec:zero-shot}, we show that with a shared prompt, the SPD has a better zero-shot transfer ability toward new languages.

\noindent \textbf{Cross-attention Decoder.} The decoder network follows a cross-attention-based multi-layer transformer architecture.
Each layer has two sub-layers. The first is a multi-head query-key-value (QKV) cross-attention module, and the second is a position-wise fully connected feed-forward network.
We employ residual connection and layer norm around each of the sub-layers.

Let $\mathbf{T}_{d_k} \in \mathbb{R}^{|d_k|\times d}$ denote the token representations generated by the encoder component for document $d_k$ in language $Y_k$, where $|d_k|$ is the number of tokens in $d_k$. 
The first decoder layer applies the cross-attention module between $\mathbf{T}_{d_k}$ and prompt matrix $\mathbf{\hat{P}}_k$. 
On the $m$th head, the attention mechanism is defined as follows:
\begin{equation*}
    \mathrm{Attention}_m = \mathrm{Softmax}\Bigl(\frac{W^q_{m}\mathbf{\hat{P}}_k \cdot W^k_{m}\mathbf{T}_{d_k}}{\sqrt{d / M}}\Bigr)W^v_{m}\mathbf{T}_{d_k}
\end{equation*}
where $M$ is the number of heads and $W^q_{m}$, $W^k_{m}$ and $W^v_{m}$ are matrices with dimension $d/M \times d$. Thus, the prompt matrix has different attention weights over encoder token representations in each subspace projection (head). 
The output of multi-head QKV cross-attention module is the concatenation of $M$ heads with linear projection:
\begin{equation*}
    \mathrm{CrossAttention}(\mathbf{\hat{P}}_k, \mathbf{T}_{d_k}) = W^o[\mathrm{Attention}_1, \ldots, \mathrm{Attention}_M]
\end{equation*}
We further define the output of the attention-based sub-layer with the residual connection and layer norm:
\begin{gather*}
\mathbf{h}_{d_k} = \mathrm{LN}\bigl(\mathbf{\hat{P}}_k + \mathrm{CrossAttention}(\mathbf{\hat{P}}_k, \mathbf{T}_{d_k})\bigr)
\end{gather*}
where $\mathrm{LN}(\cdot)$ denotes the layer norm operation. Because $\mathbf{\hat{P}}_k$ is the query element in the cross-attention module, we use the prompt matrix to query the information from the encoder output and store it in a hidden representation $\mathbf{h}_{d_k}$ which has the same dimension as the prompt matrix. Next, we apply the second sub-layer and generate the output of the first decoder layer for $d_k$, $\mathbf{H}^1_{d_k} \in \mathbb{R}^{l\times d}$: 
\begin{equation*}
    \mathbf{H}^1_{d_k} = \mathrm{DecoderLayer}_1(\mathbf{\hat{P}}_k, \mathbf{T}_{d_k}) = \mathrm{LN}(\mathbf{h}_{d_k} + \mathrm{FFN}(\mathbf{h}_{d_k}))
\end{equation*}
where $\mathrm{FFN}(\cdot)$ denotes the fully connected feed-forward network with a rectified activation function.
Then we use the hidden representation from the previous layer (i.e. $\mathbf{H}^1_{d_k}$) to query the encoder output again in the next layer, that is:
\begin{equation*}
    \mathbf{H}^{n+1}_{d_k} = \mathrm{DecoderLayer}_{n+1}(\mathbf{H}^{n}_{d_k}, \mathbf{T}_{d_k})
\end{equation*}
until reaching the maximum layer $N$ designed for the decoder. 
Finally, we average $\mathbf{H}^{N}_{d_k}$ over the prompt vector dimension as the document embedding in the target language space. A complete architecture of $f'_D$ is depicted in Figure~\ref{fig:decoder}.
\begin{equation*}
    f'_D(d_k) = \mathrm{MeanPool}(\mathbf{H}^{N}_{d_k})
\end{equation*}

\begin{figure}[t]
    \captionsetup[subfigure]{font=footnotesize,labelfont=footnotesize}
    \begin{subfigure}[t]{0.35\textwidth}
        \centering
        \includegraphics[width=\linewidth]{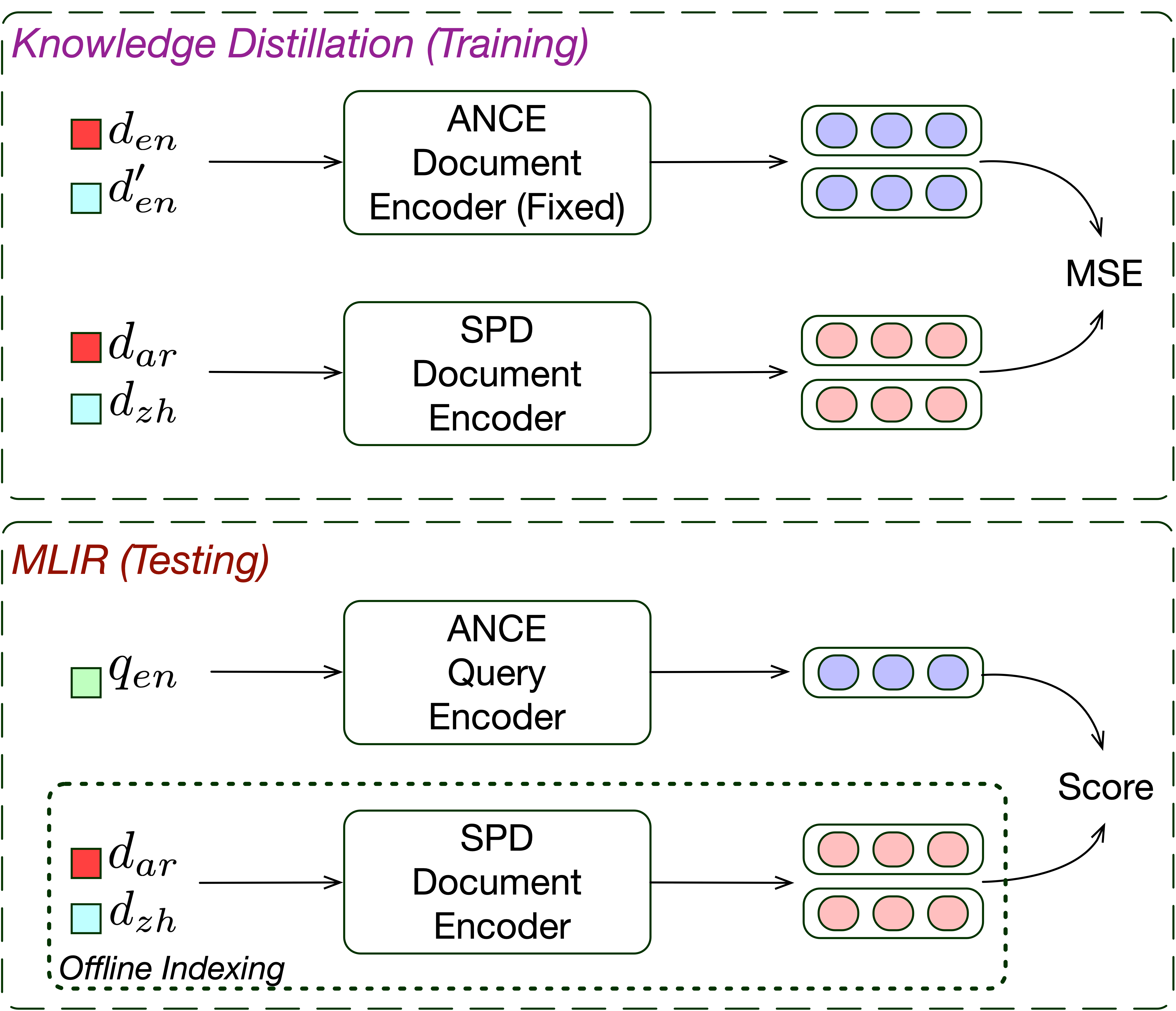}
    \end{subfigure}
\caption{Model building pipeline for \MLIR.}
\label{fig:pipeline}
\end{figure}

\subsection{Multilingual Dense Retrieval}
\noindent \textbf{Knowledge Distillation Training.} Assume that $d_{\mathrm{En}}$ is the English version of $d_k$. From the property of $g$, we know that the document embedding of $d_{\mathrm{En}}$ generated by $g_{D}$ contains rich knowledge for query-document matching in English. Equation (\ref{eq:zero_distance}) suggests that if we could let $f'_{D}$ ``behave'' like $g_{D}$, namely, if for any $d_k$, the output of $f'_D(d_k)$ is close to the output of $g_D(d_{\mathrm{En}})$, then the document embedding generated by $f'_D$ can have a similar retrieval performance as $g$ in the English domain. Therefore, we require the English document encoder $g_{D}$ as the teacher model and our multilingual document encoder $f'_{D}$ as the student. During training, we define the distillation loss as the mean square error (MSE) between two embeddings and sample $B$ examples from each language to form a batch.
\begin{equation} \label{eq:loss}
    loss := \frac{1}{KB} \sum^{K}_{k=1}\sum^{B}_{i=1} |f'_D(s_{ki}) - g_D(e_{ki})|^2
\end{equation}
where $s_{ki}$ is a sentence in language $Y_k$ and $e_{ki}$ is its parallel (translation) in English.

\noindent \textbf{Query-document matching.}
In this section, we discuss an \MLIR task of searching multilingual collections using an English query to introduce the KD-SPD framework. The query encoder in the final retrieval model can be directly copied from the teacher model in the English domain. Specifically, at test time, the matching score of $q$ and $d_*$ is calculated based on the dot-product between $g_E$ and $f'_D$:
\begin{equation*}
    f'(q, d_*) = g_E(q)f'_D(d_*)^{\top}
\end{equation*}
An overview of our \MLIR model building pipeline is shown in Figure~\ref{fig:pipeline}. 
In fact, we can also apply KD-SPD to other language settings in \MLIR task. For example, suppose the task requires searching an English collection using queries in multiple languages. In this case, KD-SPD can be built as a query encoder, and the retrieval model can reuse the teacher’s document encoder. More generally, if the \MLIR task involves a query language set $\bm{X}$ and a collection language set $\bm{Y}$, we can consider English as a bridge to build KD-SPD via two knowledge distillations: $\bm{X}$ to English for query encoder and $\bm{Y}$ to English for document encoder.


%% file: sections/experiments.tex
\section{Experimental Setup}  \label{sec:exp_setup}

\begin{table}[t]
    \centering
    \captionsetup{width=\linewidth}
    \caption{Summary of \MLIR evaluation datasets. Avg. $\#\text{d}^+$/q denotes the average number of relevant documents per query}
    \label{tab:dataset}
    \begin{adjustbox}{width=0.35\textwidth}
    \begin{tabular}{lccc}
        \toprule
        Dataset Statistics & CLEF & mTREC & LAReQA \\
        \midrule
        Query size & 133 & 54 & 1,190\\
        Collection size & 241K & 35.2M & 13,014\\
        Languages in collection & 3 & 4 & 11\\
        Avg. $\#\text{d}^+$/q  & 13.5 & 66.8 & 1.0\\
        \bottomrule
    \end{tabular}
    \end{adjustbox}
\end{table}

\subsection{Dataset}
\noindent \textbf{Evaluation data.} We focus on retrieval from multilingual collections with English queries. To comprehensively evaluate model performance on this \MLIR task, we create three test sets with various combinations of collection size, relevance distribution, and language settings. Note that some multilingual evaluation datasets have separate query sets per language, which does not thoroughly evaluate the \MLIR performance. Thus, we focus on a setting where the same set of test queries is evaluated on all languages in the collection. 
Table~\ref{tab:dataset} shows the statistics of our evaluation datasets.
\begin{itemize}[leftmargin=*,noitemsep,topsep=0pt]
\item \textbf{CLEF.} The data is from the Cross-Language Evaluation Forum (CLEF) 2000-2003 campaign for bilingual ad-hoc retrieval tracks~\cite{braschler2002clef}. We include documents in French, German, and Italian to build a multilingual collection. The English query is a concatenation of the title and description fields of the topic files. Among the CLEF C001 – C200 topics, we only consider a topic with human-annotated relevant documents in all three languages as a valid query, leading to 133 queries in total. 

\item \textbf{mTREC.} The query and relevance judgments are from the test split of the passage ranking task from the TREC 2020 Deep Learning Track~\cite{craswell2020overview}. There are three relevance judgment levels marked by {3,2,1}. We build the multilingual collection from mMARCO~\cite{bonifacio2021mmarco}, which is a machine-translated version of the MS MARCO passage collection~\cite{nguyen2016ms}. We select translated passages in four languages: Arabic, Chinese, Russian, and Indonesian, to form a large-scale multilingual collection. Because translation leads to parallel relevant documents, this evaluation set allows us to study the effect of relevant distribution over languages. We first equally distributed relevant documents on each relevance level among four languages. In section~\ref{sec:biased-dist}, we explore biased relevant distribution.

\item \textbf{LAReQA.} LAReQA~\cite{roy2020lareqa} is a benchmark for language-agnostic answer retrieval from a multilingual candidate pool. It is built based on two multilingual question-answering datasets: XQuAD~\cite{artetxe2019cross} and MLQA~\cite{lewis2019mlqa}. The query is formed using the question, and the collection is formed by breaking contextual paragraphs into sentences. Each query (question) appears in 11 different languages\footnote{Languages in LAReQA (ISO code): ar, de, el, en, es, hi, ru, th, tr, vi, zh} and has 11 parallel relevant sentences (answers). To match our \MLIR setting, we evaluate English queries on a collection of sentences in 11 languages (including English).
\end{itemize}

\noindent \textbf{Bitext training data}.
To support the multilingual knowledge distillation, we use the parallel sentences from the CCAligned dataset~\cite{elkishky_ccaligned_2020}. To train one KD-SPD model covering all three evaluation datasets (15 languages\footnote{List of training languages (ISO code): ar, de, el, en, es, fr, hi, id, it, pt, ru, th, tr, vi, zh}), we sample 4 million parallel sentences per language except English. For English, to be consistent with other languages, we sample another 4 million sentences and pair each sentence with itself.
Thus, our training data comprises 60 million sentence pairs in 15 languages. We append a language code to each sentence for SPD to identify the language of the input document.

\noindent \textbf{Retrieval fine-tuning data.}
For a competitive baseline, we further fine-tune mDPR~\cite{zhang2021mr} baseline (see section~\ref{sec:neural-baselines}) using cross-lingual triples from mMARCO~\cite{bonifacio2021mmarco}.  We sample 6 million cross-lingual triples per language to form a multilingual training set for languages in CLEF and mTREC. Because languages in LAReQA are not fully covered by mMARCO, we use mDPR on LAReQA without fine-tuning. Note that our KD-SPD model does not use this data. 

\subsection{Implementation Details}
We initialize the encoder component of the SPD model using the pre-trained XLM-R model~\cite{conneau2019unsupervised} (base-sized) and the decoder component (including prompt matrices) using the Xavier initialization~\cite{glorot2010understanding}. 
We train the SPD as a student model using bitext data. To learn the retrieval knowledge in the English domain, we employ the document encoder of ANCE~\cite{xiong2020approximate} as the teacher.  When testing, the query encoder of the final model is also a reuse of the query encoder of ANCE (except in section~\ref{sec:teacher-models}, where we investigate the impact of different teachers). For hyper-parameters, we set the length of the prompt token vector $l=30$ and the number of SPD decoding layers $N=6$. We truncate the input sequence length at 180 tokens and sample 4 examples per language to build a mini-batch. The model is trained with a learning rate of $2\times 10^{-5}$ for one epoch of all bitext data. For evaluation on the CLEF dataset, where the document length is usually longer than 180 tokens, we split long documents into overlapping passages of fixed length with a stride of 90 tokens and compute the score for each query passage pair. Finally, we select a document's maximum passage score as its ranking score~\cite{nair2022transfer}. 

\textbf{Evaluation.} We examine the top 100 ranked documents and report comprehensive metrics, including mean average precision (MAP), normalized discounted cumulative gain (nDCG@10), precision (P@10), mean reciprocal rank (MRR), and recall (R@100).
We determine statistical significance using the two-tailed paired \textit{t}-test with p-value less than 0.05 (i.e., 95\% confidence level). 

\begin{table*}[t]
    \centering
    \captionsetup{width=\linewidth}
    \caption{A comparison of model performance. The highest value is marked with bold text. For KD-SPD, statistically significant improvements are marked by $\dag$ (over mDPR) and $\ddag$ (over KD-Encoder).}
    \label{tab:main}
    \begin{adjustbox}{width=\textwidth}
    \renewcommand{\arraystretch}{1.2}
    \begin{tabular}{lccccccccccccccc}
       \toprule
       \multirow{2}{*}{\textbf{\shortstack[l]{Retrieval Method}}} & \multicolumn{5}{c}{\textbf{CLEF}} & \multicolumn{5}{c}{\textbf{mTREC}} & \multicolumn{5}{c}{\textbf{LAReQA}} \\
       \cmidrule(lr){2-6} \cmidrule(lr){7-11} \cmidrule(lr){12-16}
        & MAP & nDCG@10 & P@10 & MRR & R@100 & MAP & nDCG@10 & P@10 & MRR & R@100 & MAP & nDCG@10 & P@10 & MRR & R@100 \\
        \midrule
        SMT+Round Robin & $0.1348$ & $0.2540$ & $0.2429$ & $0.4017$ & $0.3732$ & $0.0242$ & $0.0557$ & $0.0630$ & $0.1592$ & $0.0778$ & $0.2678$ & $0.3858$ & $0.2332$ & $0.6610$ & $0.4415$\\
        SMT+Score & $0.1459$ & $0.2737$ & $0.2421$ & $0.4679$ & $0.3508$ & $0.0187$ & $0.0468$ & $0.0648$ & $0.1060$ & $0.0661$ & $0.2269$ & $0.3407$ & $0.2126$ & $0.6527$ & $0.3506$ \\
        NMT+Round Robin &$0.1783$ & $0.3732$ & $0.3474$ & $0.5793$ & $0.4118$ & $0.0653$ & $0.1735$ & $0.1870$ & $0.3965$ & $\mathbf{0.1872}$ & $0.5717$ & $0.6178$ & $0.556$ & $0.7139$ & $0.8345$ \\
        NMT+Score &$0.1950$ & $0.3806$ & $0.3474$ & $0.6140$ & $0.4206$ & $0.0522$ & $0.1570$ & $0.1685$ & $0.3970$ & $0.1691$ & $0.5063$ & $0.5671$ & $0.5178$ & $0.7091$ & $0.8002$ \\
        \midrule
        mDPR+Round Robin & $0.1823$ & $0.3412$ & $0.3165$ & $0.5448$ & $0.4330$ & $0.0490$ & $0.1358$ & $0.1537$ & $0.2913$ & $0.1324$ & $0.4935$ & $0.5223$ & $0.5163$ & $0.6493$ & $0.8394$ \\
        mDPR+Score & $0.1941$ & $0.3433$ & $0.3203$ & $0.5364$ & $0.4401$ & $0.0492$ & $0.1459$ & $0.1574$ & $0.3154$ & $0.1300$ & $0.4852$ & $0.5142$ & $0.4462$ & $0.6452$ & $0.8418$ \\
        mDPR & $0.2025$ & $0.3466$ & $0.3195$ & $0.5367$ & $0.4504$ & $0.0549$ & $0.1675$ & $0.1870$ & $0.3954$ & $0.1291$ & $0.4452$ & $0.5031$ & $0.4462$ & $0.7653$ & $0.7970$ \\
        KD-Encoder & $0.1973$ & $0.3883$ & $0.3594$ & $0.5641$ & $0.4315$ & $0.0639$ & $0.2208$ & $0.2293$ & $0.4556$ & $0.1629$ & $0.5931$ & $0.6058$ & $0.5730$ & $0.7673$ & $0.8805$ \\
        KD-SPD & $\mathbf{0.2200}^{\dag\ddag}$ & $\mathbf{0.4160}^{\dag\ddag}$ & $\mathbf{0.3714}^{\dag\ddag}$ & $\mathbf{0.6356}^{\dag\ddag}$ & $\mathbf{0.4689}^{\ddag}$ & $\mathbf{0.0748}^{\dag\ddag}$ & $\mathbf{0.2414}^{\dag\ddag}$ & $\mathbf{0.2556}^{\dag\ddag}$ & $\mathbf{0.5067}^{\dag\ddag}$ & $0.1705^{\dag}$ & $\mathbf{0.6265}^{\dag\ddag}$ & $\mathbf{0.6316}^{\dag\ddag}$ & $\mathbf{0.6049}^{\dag\ddag}$ & $\mathbf{0.7904}^{\dag\ddag}$ & $\mathbf{0.8912}^{\dag}$ \\
        \bottomrule
    \end{tabular}
    \end{adjustbox}
\end{table*}

\subsection{Compared Methods}
From a modeling perspective, we compare KD-SPD with both non-neural and neural approaches. From the system design perspective, we compare KD-SPD with end-to-end solutions and pipeline solutions via rank list merging. 

\subsubsection{Non-neural baselines}
For non-neural baselines, we generally consider a three-step pipeline to address \MLIR. First, we break the collection into subsets by language and translate the query to each subset language. Since the translated queries and subset collection are in the same language, we then use a lexical-based sparse retrieval technique (e.g., BM25) to obtain a ranked list for each language. Finally, we merge language-specific ranked lists into a final ranked list. We investigate different strategies of translation and ranked list merging that we elaborate below.

\noindent \textbf{SMT:}
We translate the query based on a statistical machine translation (SMT) method. Specifically, we first build a translation table from the parallel corpus for each language pair using GIZA++~\cite{och-ney-2003-systematic}. Then we select the top 10 translations from the translation table for each query term and apply Galago’s\footnote{https://www.lemurproject.org/galago.php/} \textit{\#combine} operator to form a translated query. Finally, we run BM25 with default parameters to retrieve documents in the same language as the query translation.

\noindent \textbf{NMT:} We translate the query into collection languages using Google Translation\footnote{https://translate.google.com/} (a neural-based commercial machine translation system). Then we run BM25 with default parameters to retrieve documents from each subset collection using the translated query.

\noindent \textbf{+Round Robin:} We merge multiple rank lists in the round-robin style, that is, iteratively extracting the top-ranked document from $K$ languages in random order to be the next $K$ of the final rank list.

\noindent \textbf{+Score:} We merge multiple rank lists by ranking scores generated by the retrieval component. Scores within each rank list are first min-max normalized to $[0,1]$. 

\noindent The non-neural baselines are the combination of translation with merging strategies: SMT+Round Robin, SMT+Score, NMT+Round Robin, and NMT+Score.

\subsubsection{Neural baselines} \label{sec:neural-baselines} As a dense retriever, we compare KD-SPD with other dense retrieval methods in the following:

\noindent \textbf{mDPR:} Models that follow the dense passage retriever (DPR) paradigm has proven to be effective for many retrieval tasks. \citet{zhang2021mr} extended DPR to non-English languages by changing the underlying pre-trained language model from BERT to multilingual BERT (mBERT). We adopt the checkpoint of mDPR trained on MS MARCO dataset~\cite{nguyen2016ms}. For CLEF and mTREC, which have fewer languages in the collections, we further fine-tune mDPR using the mMARCO dataset~\cite{bonifacio2021mmarco}. We apply mDPR to \MLIR in two ways: First, we break the \MLIR task into multiple CLIR tasks by language and use mDPR to retrieve documents from subset collections. Then we merge the rank lists from different CLIR tasks, named mDPR+Round Robin and mDPR+Score, respectively. Second, we apply mDPR as an end-to-end solution for \MLIR, in which we use it to directly index and search from the multilingual collection.

\noindent \textbf{KD-Encoder:} There are methods that can transfer the knowledge from a model built for a monolingual task to a multilingual model, enabling it to address the same task in a multilingual setting. \citet{reimers2020making} proposed a knowledge distillation method to create multilingual versions from the same monolingual models. We refer to this idea as the KD-Encoder and apply it to the \MLIR task. To compare with our approach, we adopt the same teacher model and train KD-Encoder with the same bitext data.

\section{Experimental Results}  \label{sec:exp_results}

\subsection{Retrieval Performance}
Table~\ref{tab:main} lists the evaluation results on the three \MLIR datasets. Comparing non-neural approaches, given BM25 as the same retrieval component, we can see that methods based on NMT outperform those based on SMT. For document collections with mostly high-resource languages, NMT based method can also achieve higher nDCG, precision, and MRR scores than end-to-end neural approaches (i.e., NMT+Score on CLEF). It highlights that translation quality is an important factor in \MLIR.

Usually, for a pipeline approach, the error can accumulate for each step and lead to a sub-optimal result~\cite{ferreira2019neural, glasmachers2017limits}. In \MLIR, without evaluating the content with respect to the query, merging rank lists only based on the score or rank within sub-collection will cause errors from multiple languages to accumulate.
However, comparing the pipeline with the end-to-end approach of mDPR, we can see that end-to-end mDPR does not show a consistent advantage over the pipeline mDPR. There are two plausible reasons.
First, similar to other multilingual models, mDPR based on a multilingual pre-trained language model also inherits the language bias in the pre-training step. Second, the fine-tuning steps of mDPR only focus on ranking documents within the same language space. Even trained with multilingual retrieval data, the candidate documents are still monolingual, and the score comparison is between two particular languages.
These two reasons cause the ranking score generated by mDPR to be inconsistent across languages.
Moreover, KD-Encoder performs better than mDPR on mTREC and LAReQA, On CLEF, it also scores higher nDCG, precision, and MRR than mDPR. Such results suggest that mapping parallel text from different languages to the same location in the vector space via knowledge distillation can efficiently transfer monolingual retrieval knowledge to multilingual settings. Finally, with the support of soft prompt decoding, KD-SPD achieves the best retrieval performance among all compared methods. In terms of precision-oriented metrics, it consistently and significantly outperforms both mDPR and KD-Encoder.

\subsection{Biased Relevant Distribution} \label{sec:biased-dist}

\begin{table}[t]
    \centering
    \captionsetup{width=\linewidth}
    \caption{Performance comparison of biased distributed relevant documents in mTREC. Significance tests are marked by $\dag$ (over mDPR) and $\ddag$ (over KD-Encoder).}
    \label{tab:biased}
    \begin{adjustbox}{width=0.42\textwidth}
    \renewcommand{\arraystretch}{1.2}
    \begin{tabular}{lccccc}
       \toprule
       \multirow{2}{*}{\textbf{\shortstack[l]{Retrieval Method}}} & \multicolumn{5}{c}{\textbf{Biased mTREC}} \\
       \cmidrule(lr){2-6}
        & MAP & nDCG@10 & P@10 & MRR & R@100 \\
        \midrule
        SMT+Round Robin & $0.0134$ & $0.0304$ & $0.0426$ & $0.0759$ & $0.0621$ \\
        SMT+Score & $0.0149$ & $0.0356$ & $0.0426$ & $0.1083$ & $0.0698$ \\
        NMT+Round Robin & $0.0331$ & $0.0939$ & $0.1278$ & $0.2902$ & $0.1500$ \\
        NMT+Score & $0.0438$ & $0.1055$ & $0.1389$ & $0.3430$ & $\mathbf{0.1751}$ \\
        \midrule
        mDPR+Round Robin & $0.0301$ & $0.0902$ & $0.1074$ & $0.2922$ & $0.1150$ \\
        mDPR+Score & $0.0516$ & $0.1576$ & $0.1778$ & $0.3655$ & $0.1206$ \\
        mDPR & $0.0508$ & $0.1571$ & $0.1759$ & $0.3652$ & $0.1174$ \\
        KD-Encoder & $0.0681$ & $0.2028$ & $0.2078$ & $0.4055$ & $0.1494$ \\
        KD-SPD & $\mathbf{0.0753}^{\dag}$ & $\mathbf{0.2317}^{\dag\ddag}$ & $\mathbf{0.2352}^{\dag\ddag}$ & $\mathbf{0.4579}^{\dag\ddag}$ & $0.1684^{\dag\ddag}$ \\
       \bottomrule
    \end{tabular}
    \end{adjustbox}
\end{table}

In the \MLIR task, some queries strongly prefer one language over others and some do not. Thus, different queries tend to have different relevant document distributions among languages. This special feature requires the retrieval system to rank documents independent of their language.
In the experiment on mTREC shown in Table~\ref{tab:main}, relevant documents were distributed equally among four languages for each query. The parallel translations of mTREC allow us to test with different relevant document distributions. In this section, we simulate the language preference in \MLIR task: For each query, we first randomly select a language as the primary language and assign 60\% of the top relevant documents (sorted by relevance judgment level) to that language. And the other three become the minor languages for this query, among which we equally distribute the remaining 40\% of the relevant documents. Table~\ref{tab:biased} shows the results on biased distributed relevant documents of mTREC. As expected, the performance of methods based on round-robin merge drop significantly. The reason is that the rank list from minor languages introduces more errors compared to the scenario where languages are uniformly distributed. We can see that KD-SPD is also affected by the change in distribution yet still performs the best among all compared methods.

\subsection{Analysis of Knowledge Distillation}
\begin{figure}[t]
    \captionsetup[subfigure]{font=footnotesize,labelfont=footnotesize}
    \begin{subfigure}[t]{0.48\textwidth}
        \centering
        \includegraphics[width=\linewidth]{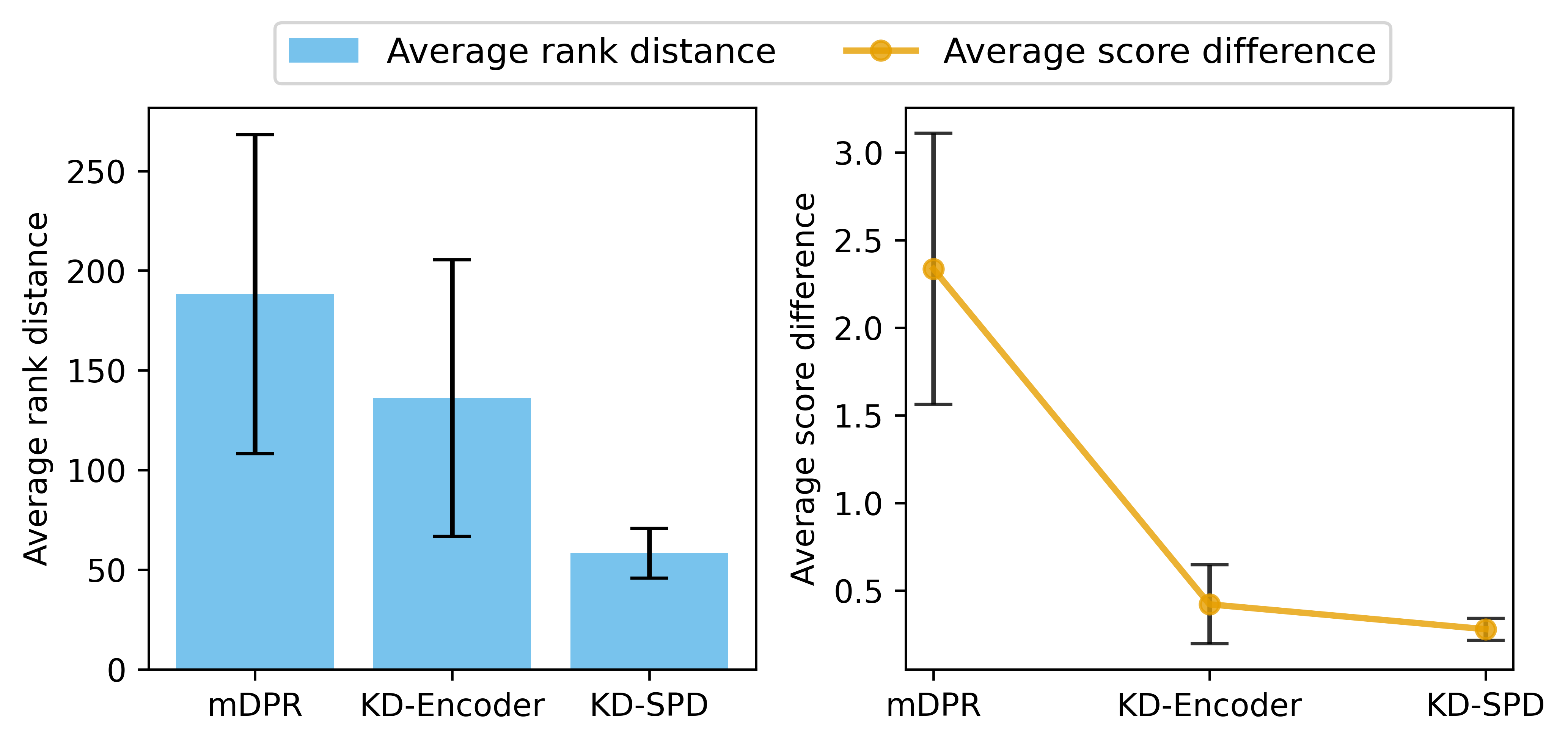}
    \end{subfigure}
\vspace{-18pt}
\caption{Parallel document analysis for \MLIR models.}
\label{fig:rank-distance}
\end{figure}

To study how SPD behaves after knowledge distillation, we compare the rank distance and score difference of parallel relevant documents in the rank lists generated by different models. 
In this experiment, again, we take advantage of parallel translations in mTREC and build \textit{duplicate} relevant documents in four languages. Thus, for each query, there are semantically similar relevant documents in different languages.
Given a query, we locate all parallel relevant documents in four languages within the top 1,000 candidates from rank lists generated by mDPR, KD-Encoder, and KD-SPD, respectively.
Then we compute the maximum rank distance and score difference among the four parallel documents. 
The equation to compute the score difference is as follows:
\begin{equation*}
\mathcal{S} = \frac{1}{|\mathcal{Q}|}\displaystyle\sum_{i=1}^{|\mathcal{Q}|} \frac{1}{|\mathcal{R}_{q_i}|} \displaystyle\sum_{d_{kj} \in \mathcal{R}_{q_i}} \Big( \max_{k \in \bm{Y}}f' (q_i, d_{kj}) - \min_{k \in \bm{Y}} f' (q_i, d_{kj}) \Big)
\end{equation*}
where $\mathcal{Q}$ is the query set, $\mathcal{R}_{q_i}$ is the set of relevant documents for the query $q_i$, and $\bm{Y}$ is the language set. The averaging rank distance can also be obtained in a similar way.
Figure~\ref{fig:rank-distance} shows the results averaged over 54 queries in mTREC. We can see that KD-SPD has the smallest rank distance and score difference over parallel documents. 
The rank and score of parallel documents reflect the language bias in \MLIR models. Thus, KD-SPD is less biased toward languages when ranking documents from a multilingual collection. Moreover, because the query embedding is fixed given the same query, the low mean and standard deviation values indicate that KD-SPD is able to generate similar embeddings for parallel documents in different languages. This matches the model design purpose.

\subsection{Ablation Study}
\label{sec:teacher-models}

\begin{table*}[t]
    \centering
    \captionsetup{width=\linewidth}
    \caption{Ablation I: Decoder architecture. The numbers in the bracket are differences in percentage to KD-Encoder.}
    \label{tab:ablation-size}
    \begin{adjustbox}{width=0.95\textwidth}
    \renewcommand{\arraystretch}{1.2}
    \begin{tabular}{lccccccccccc}
       \toprule
       \multirow{2}{*}{\textbf{\shortstack[l]{Model}}} & \multirow{2}{*}{\textbf{\shortstack[c]{Parameter\\Size}}} & \multicolumn{5}{c}{\textbf{CLEF}} & \multicolumn{5}{c}{\textbf{LAReQA}}\\
       \cmidrule(lr){3-7} \cmidrule(lr){8-12}
        &  & MAP & nDCG@10 & P@10 & MRR & R@100 & MAP & nDCG@10 & P@10 & MRR & R@100 \\
        \midrule
        KD-Encoder & 278.6M & 0.1973 & 0.3883 & 0.3594 & 0.5641 & 0.4315 & 0.5931 & 0.6058 & 0.5730 & 0.7673 & 0.8805 \\
        KD-SPD & 320.0M (+14.8) & 0.2200 (+11.5) & 0.4160 (+7.1) & 0.3714 (+3.3) & 0.6356 (+12.7) & 0.4689 (+8.7) & 0.6265 (+5.6) & 0.6316 (+4.2) & 0.6049 (+5.6) & 0.7904 (+3.0) & 0.8912 (+1.2) \\
        KD-UTSPD & 284.5M (+2.1) & 0.2075 (+5.2) & 0.4023 (+3.6) & 0.3722 (+3.6) & 0.5964 (+5.7) & 0.4576 (+6.0) & 0.6212 (+4.7) & 0.6279 (+3.6) & 0.5996 (+4.6) & 0.7674 (+0.0) & 0.8870 (+0.7)\\
       \bottomrule
    \end{tabular}
    \end{adjustbox}
\end{table*}

\begin{table*}[t]
    \centering
    \captionsetup{width=\linewidth}
    \caption{Ablation II: Effect of Teacher model. Significance tests with respect to KD-SPD (ANCE) are marked in $\blacktriangle$.}
    \label{tab:ablation-teacher}
    \begin{adjustbox}{width=0.7\textwidth}
    \renewcommand{\arraystretch}{1.2}
    \begin{tabular}{lcccccccccc}
       \toprule
       \multirow{2}{*}{\textbf{\shortstack[l]{Teacher}}} & \multicolumn{5}{c}{\textbf{CLEF}} & \multicolumn{5}{c}{\textbf{LAReQA}}\\
       \cmidrule(lr){2-6} \cmidrule(lr){7-11}
        &  MAP & nDCG@10 & P@10 & MRR & R@100 & MAP & nDCG@10 & P@10 & MRR & R@100 \\
        \midrule
        KD-SPD (ANCE) & 0.2200 & 0.4160 & 0.3714 & 0.6356 & 0.4689 & 0.6265 & 0.6316 & 0.6049 & 0.7904 & 0.8912 \\ 
        KD-SPD (coCondenser) & $0.2487^{\blacktriangle}$ & $0.4546^{\blacktriangle}$ & $0.4008^{\blacktriangle}$ & $0.6826^{\blacktriangle}$ & $0.4976^{\blacktriangle}$ & $0.6501^{\blacktriangle}$ & $0.6694^{\blacktriangle}$ & $0.6436^{\blacktriangle}$ & 0.8012 & 0.9172 \\
       \bottomrule
    \end{tabular}
    \end{adjustbox}
\end{table*}

In this section, we conduct experiments on two aspects that could affect the performance of KD-SPD: The number of layers in the decoder and the choice of the teacher model for distillation.

\noindent\textbf{Decoder architecture.}
Following the idea of weights share in Transformers~\cite{dehghani2018universal, jaegle2021perceiver}, we replace the multi-layer (6-layer) decoder with a recurrent decoder block. Instead of $N$ distinct layers, a decoder block has the same architecture as one decoder layer and is called recurrently for $N=12$ steps. The weights of a decoder block are shared between steps. After each step, we add a temporal embedding $\boldsymbol\tau \in \mathbb{R}^{l\times d}$ to the hidden states. 
\begin{equation*}
    \mathbf{H}^{n+1}_{d_k} = \boldsymbol{\tau}_n + \mathrm{DecoderBlock}(\mathbf{H}^{n}_{d_k}, \mathbf{T}_{d_k})
\end{equation*}
This approach significantly reduces the size of model parameters. Named universal transformer-based SPD (UTSPD), Table ~\ref{tab:ablation-size} shows its performance, compared to KD-Encoder and KD-SPD. We can see that only with 2.1\% more parameters, KD-UTSPD performs better than KD-Encoder. By reducing the parameter size, we show that the performance gain in SPD mainly relies on the prompt design and decoder component based on the cross-attention module. Because reducing parameters limits the model’s generalization ability, there is a performance drop from distinct layers to shared weights.

\noindent\textbf{Teacher model}
The teacher model bounds the retrieval performance of KD-SPD. We hypothesize that a better teacher model in the English domain can lead to a better SPD model for \MLIR task. Based on the leaderboard of MS MARCO passage ranking, we replace ANCE~\cite{xiong2020approximate} with coCondenser~\cite{gao2021unsupervised} for knowledge distillation. To be consistent with coCondenser, we also change the pre-trained multilingual language model used in SPD from XLM-R to mBERT. 
The evaluation of SPD trained with different teacher models is shown in Table~\ref{tab:ablation-teacher}. In general, KD-SPD learned from coCondenser performs better than the one learned from ANCE. 
This suggests that improvements with respect to the retrieval performance in the English domain can be transferred to \MLIR task via KD-SPD.

\subsection{Zero-shot Transfer}
\label{sec:zero-shot}
\begin{table}[t]
    \centering
    \captionsetup{width=\linewidth}
    \caption{Zero-shot CLIR: English-to-Finnish. Significance tests are marked by $\dag$ (over mDPR) and $\ddag$ (over KD-Encoder).}
    \label{tab:zero-finnish}
    \begin{adjustbox}{width=0.38\textwidth}
    \renewcommand{\arraystretch}{1.2}
    \begin{tabular}{lccccc}
       \toprule
       \multirow{2}{*}[-4pt]{\textbf{\shortstack[l]{Retrieval\\Method}}} & \multicolumn{5}{c}{\textbf{CLEF Finnish}} \\
       \cmidrule(lr){2-6}
        & MAP & nDCG@10 & P@10 & MRR & R@100 \\
        \midrule
        SMT & 0.0739 & 0.1179 & 0.0900 & 0.1390 & 0.1828 \\
        NMT & 0.1613 & 0.2562 & 0.1560 & 0.4591 & 0.4251 \\
        \midrule
        mDPR & 0.1682 & 0.2143 & 0.1300 & 0.3095 & 0.5010 \\
        KD-Encoder & 0.1845 & 0.2796 & 0.1920 & 0.4537 & 0.5237 \\
        KD-SPD & \textbf{0.2286}$^{\dag\ddag}$ & \textbf{0.3321}$^{\dag\ddag}$ & \textbf{0.2220}$^{\dag\ddag}$ & \textbf{0.5092}$^{\dag\ddag}$ & \textbf{0.5958}$^{\dag\ddag}$ \\
       \bottomrule
    \end{tabular}
    \end{adjustbox}
\end{table}

\begin{table}[t]
    \centering
    \captionsetup{width=\linewidth}
    \caption{Zero-shot \MLIR. Significance tests are marked by $\dag$ (over mDPR) and $\ddag$ (over KD-Encoder).}
    \label{tab:zero-mlir}
    \begin{adjustbox}{width=0.42\textwidth}
    \renewcommand{\arraystretch}{1.2}
    \begin{tabular}{lccccc}
       \toprule
       \multirow{2}{*}[-4pt]{\textbf{\shortstack[l]{Retrieval\\Method}}} & \multicolumn{5}{c}{\textbf{CLEF DE-IT-FI}} \\
       \cmidrule(lr){2-6}
        & MAP & nDCG@10 & P@10 & MRR & R@100 \\
        \midrule
        SMT+Round Robin & 0.1099 & 0.2245 & 0.208 & 0.4096 & 0.2909 \\
        SMT+Score & 0.1269 & 0.2242 & 0.218 & 0.3726 & 0.2974 \\
        NMT+Round Robin & 0.1263 & 0.2748 & 0.254 & 0.5039 & 0.3384 \\
        NMT+Score & 0.1447 & 0.2806 & 0.258 & 0.5101 & 0.344 \\
        \midrule
        mDPR+Round Robin & 0.1481 & 0.2734 & 0.268 & 0.391 & 0.3974 \\
        mDPR+Score & 0.1728 & 0.3002 & 0.282 & 0.4816 & 0.4083 \\
        mDPR & 0.1952 & 0.3377 & 0.306 & 0.5175 & 0.4107 \\
        KD-Encoder & 0.1963 & 0.4262 & 0.382 & 0.6753 & 0.4152 \\
        KD-SPD & \textbf{0.2174}$^{\dag\ddag}$ & \textbf{0.4494}$^{\dag\ddag}$ & \textbf{0.4100}$^{\dag\ddag}$ & \textbf{0.7099}$^{\dag\ddag}$ & \textbf{0.4545}$^{\dag\ddag}$ \\
       \bottomrule
    \end{tabular}
    \end{adjustbox}
\end{table}

We explore the zero-shot ability of KD-SPD. For documents in languages that are not observed in the training data, we first define the language-specific vectors by averaging all trained language-specific vectors from known languages. 
Then KD-SPD follows the same steps as other languages to generate a prompt matrix for the new language. 
Hence, observed languages' knowledge transfers to the new language via the shared prompt matrix.

In this study, we focus on Finnish as the target language and use a collection of 54,694 Finnish documents from the CLEF dataset. It's worth mentioning that Finnish, a member of the Uralic language family, is distinct from the 15 languages used in training. Among the 133 English queries in the CLEF dataset, 50 have relevant annotations in the Finnish collection, forming a new set of test queries. The results in Table 6 show the performance of KD-SPD in Cross-Language Information Retrieval (CLIR) between English and Finnish, and we observe that KD-SPD significantly outperforms other methods, demonstrating the transferability of knowledge from the prompt matrices to new languages.
Next, we expand the evaluation to a more challenging setting, combining Finnish with German and Italian. The resulting collection contains both observed and unobserved languages. Table 7 shows KD-SPD's zero-shot performance in the multilingual information retrieval (MLIR) setting, where it still achieves the best results. This highlights KD-SPD's strong ability to transfer knowledge in a zero-shot scenario.

%% file: sections/conclusion.tex
\section{Conclusions and Future Work}  \label{sec:conclusion}
In this work, we presented a knowledge distillation (KD) framework based on soft prompt decoding (SPD) to address the multilingual information retrieval (\MLIR) task. Using the soft prompt matrix as a task indicator, KD-SPD can implicitly translate documents from multiple languages into the same embedding space as the query language. We proposed prompt decomposition to enable efficient knowledge sharing across all source languages.
Our knowledge distillation framework transfers knowledge from a well-trained monolingual retrieval model to KD-SPD, greatly reducing the retrieval data requirements for training \MLIR models. 
Our comprehensive experimental results show that KD-SPD significantly outperforms other baselines on three qualitatively different \MLIR evaluation datasets. Further analysis demonstrates that KD-SPD has less language bias and better zero-shot transfer ability toward new languages.
For future work, as a general knowledge transfer framework, we are interested in extending KD-SPD to transfer other monolingual task-specific knowledge into the multilingual space. Exploring the applications of KD-SPD to multimodal information retrieval is also an exciting future direction.